\documentclass[11pt]{amsart}
\usepackage[utf8]{inputenc}
\usepackage[english]{babel}
\usepackage{amsmath}
\usepackage{amsfonts}
\usepackage{amssymb}
\usepackage[all]{xy}
\usepackage{graphicx}
\usepackage{xcolor}
\usepackage{hyperref}
\usepackage{tabularx}

\theoremstyle{definition}
\newtheorem{example}{Example}[section]
\newtheorem{remark}{Remark}[section]

\begin{document}

\title{Modelling Stochastic Time Delay for Regression}
\author[Aaron Pickering]{Aaron Pickering}
\email{\href{mailto:aaron_rcl@hotmail.com}{aaron\_rcl@hotmail.com}}
\author[Juan Camilo Orduz]{Juan Camilo Orduz}
\email{\href{mailto:juanitorduz@gmail.com}{juanitorduz@gmail.com}}
\urladdr{\href{https://juanitorduz.github.io/}{juanitorduz.github.io}}

\date{\today}

\begin{abstract}
Systems with stochastic time delay between the input and output present a number of unique challenges. 
Time domain noise leads to irregular alignments, obfuscates relationships and attenuates inferred coefficients. 
To handle these challenges, we introduce a maximum likelihood regression model that regards stochastic time delay as an 'error' in the time domain. For a certain subset of problems, by modelling both prediction \emph{and} time errors it is possible to outperform traditional models.
Through a simulated experiment of a univariate problem, we demonstrate results that significantly improve upon Ordinary Least Squares (OLS) regression.
\end{abstract}

\maketitle

\section{Introduction}

Consider the typical univariate regression problem where the intention is to find the relationship between $x$ and $y$. When extended to time series, a number of time specific complexities arise. For the specific case where the input $x$ affects the output $y$, with a random time delay in between, the estimated coefficients (or weights, predictions etc) are significantly attenuated \cite{spearman}. This attenuation occurs both in traditional statistical models and machine learning models, and for certain problems can be a serious limitation. The following document proposes techniques for handling this class of time series regression.

As an example, let us take the management of blood sugar level in diabetes patients. In people with diabetes, the pancreas can't effectively regulate blood sugar levels. Therefore, these levels must be controlled by insulin injections and a special diet. The challenge for many people is that the relationship between the input (insulin) and output (blood sugar) is extremely complex \cite{blood_glucose}. The effect of the insulin injection might be observed after $15$, $20$  or $t$ minutes depending on a number of factors, many of which are unknown. Due to the stochastic nature of the time delays, the actual effect can't be easily determined. It is difficult to differentiate the effect of the insulin injection from other factors and accurately determine how much one should take.

Inference for this type of problem is especially challenging. Typical regression models require a fixed alignment between cause and effect. Using standard methods, we'd need to assume that the effect occurs after some fixed time $t$ which can be inferred from the data. However, if there is any uncertainty in the parameter $t$ ($t$ changes or is noisy) the resulting estimates will be significantly attenuated.

Consider the simple example in Figure 1 where the effect of the input is $1$. The observed input is given by the red line, the blue line is when the effect actually occurs. The first effect happens one time point after the input. The second effect happens at the same time as the input. A fixed time delay isn't valid in this case because the time shifts differ.

\begin{center}
\begin{figure}
\includegraphics[totalheight=6cm]{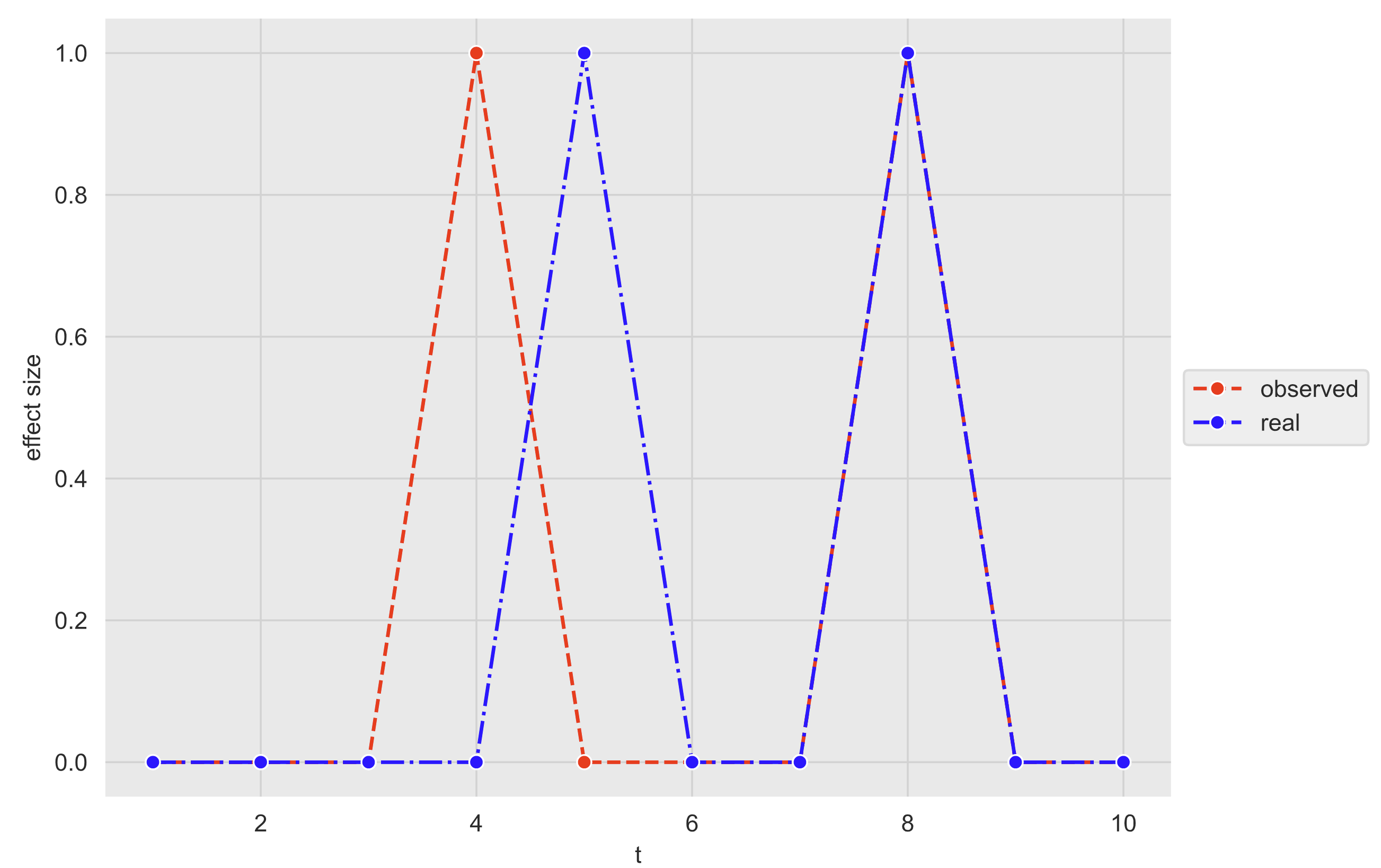}
\caption{Stochastic Time Delay Example}
\label{fig:verticalcell}
\end{figure}
\end{center}

If one were to model the effect using a fixed time delay and OLS, the estimate would only be half the true value because only one of the outputs is aligned. Obviously, this isn't ideal, we want the parameter estimates to be as close to the real values as possible, regardless of any noise in the lag structures. One can mitigate the problem via time aggregations, however in complicated cases with multiple factors, this is just not feasible.

Therefore, we propose a regression model which can handle stochastic time delay structures. We treat the stochastic time delay components as an ‘error’ in the time axis. Next, we find the maximum likelihood estimate, for a given set of parameters, considering the time error ($t$-axis) and the regression error ($y$-axis) simultaneously.\\

An open source Python implementation is available on \href{https://github.com/aaron1rcl/tvs_regression/}{GitHub}.

\section{Related Work}
There is extensive existing literature on time series problems with time delay dependencies. In the statistical disciplines there a number of treatments which focus on 
fixed time delay dependencies. For example, Granger Causality \cite{granger} is used to determine whether one time series is useful for forecasting another, across some fixed delay.
Distributed Lag \cite{almon} and Dynamic Regression models (e.g \cite{dynr}) are able to handle linear and non-linear cause and effect relations that occur across multiple time lags.
Nonetheless, these models assume a fixed time lag dependency.

In the machine learning literature, there are a number of models which can handle complex dependencies across time. Sequence models such as the RNN, LSTM and GRU \cite{ml}
are able to generate predictions which incorporate time delays between input and the output variables. More recently, attention based models such as the Transformer have become the state-of-the-art for a
variety of tasks, including time series forecasting \cite{deep_transformers}. Yet, all of these models have a tacit assumption of a fixed time lag dependency. There is also Dynamic Time Warping (DTW) \cite{dtw} which tries to find the optimal alignment between time series sequences by minimizing the
distance between the respective inputs. DTW is able to model varying time lag dependencies but is not purposed for regression, instead being mainly used for pattern matching and sequence alignment.

The field of system identification also has considerable literature on dynamic non-linear systems. For example, the NARX \cite{ml} model can be used to identify 
non-linear systems with fixed time delays between input and output. At the same time, there are also treatments on systems with disturbances in the input \cite{eiv_sysid}, known as EIV systems. To our knowledge,
these works do not deal with uncertainty in the time domain.

Finally, recent papers such as Dynamic Time Lag Regression \cite{dtlr} and Variable-Lag Granger Causality \cite{Amornbunchornvej_2019} deal with non-stationary time lag dependencies. In other words, the lag structure is assumed to evolve over time.

In this paper, we deal with the specific case of \emph{stochastic time delays} i.e. time delays which vary randomly. There appears to be little existing research on this topic.

\section{Methodology}

We consider the problem as analogous to the typical error-in-variables (EIV) regression. Ordinary regression analyses (and machine learning models) define the loss function with respect to errors in the $y$ axis only. For EIV, errors are considered in both the $y$-axis and the $x$-axis \cite{deming}. This is useful when there are measurement errors in the independent variable e.g. because the physical measurements have some degree of random error. Similarly, for this problem we assume that we have errors in the $y$-axis and the $t$-axis. That is, there are random prediction errors and random errors in the time domain. 

\subsection{Model Specification}

Let $x(t)$ be a discrete time series. We want to determine the functional relationship $f:\mathbb{R} \longrightarrow \mathbb{R}$ to some other (target) variable $y$ on observed data $\{y_i\}_{i=1}^{n}$, that is
$y = f(x(t))$.  The series $x(t)$ must be stationary with a known value for the point at which $f(x(t)) = 0$.  That is, we know at which point the input series has no effect on the output (when it’s off). The size of the support of $f(x(t))$ should be small relative to number of points in the domain (we we will explore this condition further in 'Limitations'). We utilise the terminology {\em impulse} for an individual element of the support of $f$.

Firstly, let us take the input series $x(t)$ and decompose it into its constituent non-zero impulse components. So, if $x(t)$ is a vector given by 

\begin{align*}
x(t) = 
\left(
\begin{array}{cccccc}
0 & 0 & 1 & 0 & 1 & 0
\end{array}  
\right)
\end{align*}

then we decompose the vector into a matrix 

\begin{align*}
X(t) = 
\left(
\begin{array}{cccccc}
0 & 0 & 1 & 0 & 0 & 0 \\
0 & 0 & 0 & 0 & 1 & 0 
\end{array}  
\right)
\end{align*}

where each impulse is treated separately. \\

Given that each non-zero impulse (row) of $X(t)$ is affected by a random time delay (denoted $\tau$), we then model:

\begin{align}\label{Eqn:ModelSpecification}
y(t)= f(1^TX(t + \mathcal{T})) + \varepsilon
\end{align}

where $X(t + \mathcal{T})$ is the matrix of time shifted impulses, $X(t)$ is the observed input, $\mathcal{T}$ is the set of single impulse time delays ($\tau$'s), $1$ is the vector $(1,1,\cdots, 1)\in 
\mathbb{R}^{1 \times k}$, where $k$ is the number of impulses and $\varepsilon$ is a noise term. 

The model definition represents the application of time shifts to the matrix rows and a subsequent reduction by summation over the columns. 

We also assume that $\mathcal{T}$ is not a constant, but rather a random draw from some discrete distribution (e.g. discrete gaussian, poisson etc). For the discrete gaussian kernel $\mathcal{T}\sim T(\mu, \sigma)$ or for the poisson distribution $\mathcal{T}\sim Pois(\lambda)$. Similarly one can model the errors as  $\varepsilon \sim N(\mu, \sigma_{\varepsilon})$. 
\subsection{Inference}

In order to find the best estimate of $f$, we would like to find the function $f$ which maximises the joint log-likelihood of the time-domain shifts and the prediction residuals. Specifically, we maximise:

\begin{align}
    \mathcal{L} = 
    \underbrace{\sum_{i=1}^{n} \log(p(y_i | X_{\tau_i} = x_i; \theta_{f}, \tau_i))}_{\mathcal{L}_1} 
    + 
    \underbrace{\sum_{i=1}^{n} \log(p(\tau_i| \theta_{\mathcal{T}}))}_{\mathcal{L}_2}
\end{align}

where $\theta_{f}$ and $\theta_{\mathcal{T}}$ represent the parameters of the model $f$ and time shifts $\mathcal{T}$ respectively. In other words, the $\mathcal{L}_2$ term represents likelihood of time shifts and the $\mathcal{L}_1$ term represents the likelihood of the prediction residuals. We maximise these terms simultaneously. For simplicity, we assume that the time shift distribution and error distributions are independent. \\

We will refer to this algorithm as {\em Time Varying Stochastic (TVS) Regression}.

\section{Algorithm}

Before optimisation, the values of each individual time shift $\tau$ are not known. In addition, the prediction errors $\varepsilon$ can only be calculated if each value of $\tau$ is available (because for each time shift there is a different prediction and hence prediction error). Therefore, our algorithm finds the optimum set of time shifts in an inner optimisation ($\mathcal{L}_2$), while iteratively searching for the optimum parameters of $f$ in an outer optimisation loop ($\mathcal{L}_1$). 
Firstly, we define a parametric form $f(x;\theta_f)$ and some initial parameters to be estimated for the function $f$. 

\begin{example}\label{example:poisson}
Let's take the univariate linear model with gaussian errors, where the $f(x(t))$ is parameterised by $\beta$ and $\sigma_{\varepsilon}$ (error standard deviation). In addition, we choose a poisson distribution for $\mathcal{T}$, where $\theta_{\mathcal{T}}$ is a parameterised by its mean $\lambda_{\mathcal{T}}$. The choice of a poisson distribution for $\mathcal{T}$
ensures discrete, positive time shifts only.
\end{example}

First, let us initialise some starting values for each of these parameters. Now, we want to find the best possible time shift $\tau$ for each input impulse in $X(t)$. It stands to reason that the best possible time shift would be one that is not too distant from the observed impulse (i.e. has a high likelihood given some distribution) and also produces the best possible prediction. From there, the likelihood estimate is derived from the time shift and the prediction error.
Finally, we iterate over a number of values of $\tau$ optimising until we maximise the likelihood for the specific impulse.

\begin{remark}
However, we must also consider that the impulses in $X(t)$ are not independent from each other. After shifting, it's possible that two or more effects occur simultaneously. This is particularly problematic if there are multiple impulses within a short period of time or impulses have a distributed effect over multiple time points. As an example consider the series 

\begin{align*}
x(t) = 
\left(
\begin{array}{cccccc}
0 & 0 & 1 & 1 & 0 & 0
\end{array}  
\right)
\end{align*}

with $\mathcal{T}= (1, 0)$ and $\beta = 1$. For this case, 

\begin{align*}
X(t + \mathcal{T}) = 
\left(
\begin{array}{cccccc}
0 & 0 & 1 & 0 & 0 & 0 \\
0 & 0 & 1 & 0 & 0 & 0 
\end{array}  
\right)
\end{align*}

and the effect is therefore 
\begin{align*}
y = 
\left(
\begin{array}{cccccc}
0 & 0 & 0 & 2 & 0 & 0
\end{array}  
\right)
\end{align*}
\end{remark}

To accurately calculate the likelihood, we must optimise the time shifts simultaneously. Therefore, we treat the problem of finding the best set of time shifts as a discrete optimisation problem. For the optimisation step we utilise two assumptions.

\begin{enumerate}
\item Firstly, smaller shifts are more likely than larger shifts (proportionate to the dispersion of the $\mathcal{T}$ distribution). The algorithm should explore the space of smaller shifts more often than larger shifts. 
\item Secondly, impulses close to each other are more likely to be dependent than impulses further away. 
\end{enumerate}

Accordingly, the optimisation procedure is:

\begin{enumerate}
    \item [(1)] Initialise the set of parameters $\theta_{f}$ and $\theta_{\mathcal{T}}$ for the function $f$.
    \item [(2)] Find the $\mathcal{T}$ which maximises the likelihood for the given set of parameters:
        \begin{enumerate} 
            \item[(i)] Initialize the discrete optimisation algorithm time shifts ($\mathcal{T}$). 
            \item[(ii)] Next, randomly select a small number of impulses $m$, with the value of $m$ treated as a hyperparameter. For each impulse, a random time shift is drawn from the $\mathcal{T}$ distribution creating a proposal vector. 
            \item [(iii)]The likelihood (both time shift and prediction error) for the proposal is calculated. 
            \item[(iv)] If the proposal likelihood is higher than the current maximum likelihood, our estimate is updated.
            \item [(v)] Return to (i) and repeat. The best estimate of the set of $\mathcal{T}$ improves each iteration. The number of iterations ($N$) is also treated as a hyperparameter.
        \end{enumerate} 
    \item [(3)] Optimise the model parameters $\theta_{f}$ and $\theta_{\mathcal{T}}$.
\end{enumerate}

\begin{remark}
In Example \ref{example:poisson} we recommend to initialize the discrete optimisation algorithm with all parameters set to the mean of the $\mathcal{T}$ distribution ($\lambda_{\tau}$).
\end{remark}

For the outer parameter optimisation (Step (3)), typical methods such as gradient descent, genetic algorithms and simulated annealing can be used.  In our implementation, we have used the differential evolution algorithm \cite{diff_ev} (scipiy.optimize \cite{scipy}) because of its ability to handle noisy objective functions \cite{diff_ev}. In order to improve convergence, we also standardize all input variables to the range (0,1).

\begin{remark}
We also note that the accuracy of the final parameter estimate is relative to the ratio of the $y$-axis error and the effect size $f(x(t))$. 
As the ratio of noise $\sigma_{\varepsilon}$ to effect $f(x(t))$ increases, the time shift distribution $\mathcal{T}$ shrinks. 
In the event that the mean of $\theta_{\mathcal{T}}$ become 0, the model becomes a ‘fixed lag’ model, and the parameter estimate $\beta$ tends to the standard linear regression coefficient. Therefore, the method provides no guarantee on recovering the exact time shifts, only that the coefficient estimates are equal to or better than their OLS counterparts.
\end{remark}

\section{Limitations}
\subsection{Scaling}
As the length of $x(t)$ increases, more impulses are introduced and the size of the decomposed matrix $X(t)$ also increases. At some point, handling $X(t)$ becomes impractical. To account for this, we assume that distant impulses do not affect each other. Concretely, if the $i_{th}$ and $j_{th}$ impulse (row) in $X(t)$ are far away then their effect vectors are orthogonal. Hence, we can decompose the matrix $X(t)$ into orthogonal blocks on which the inner optimisation can be run in parallel.

\subsection{Constraints}
Finally, a note on problem constraints. The likelihood estimates are dependent on the inner optimisation procedure. There is no guarantee that the global maximum will be found, particularly for sequences with a high density of impulses. 
When the density (in time) of non-zero impulses in $x(t)$ is high, the possible solution space of $\mathcal{T}$ grows exponentially.
In such cases, the estimated likelihood is likely to be close to, but not exactly the same as the real value. Therefore, TVS Regression is most appropriate for sparse time series inputs.

\section{Experiment}

\begin{center}
\begin{figure}
\includegraphics[scale=0.5]{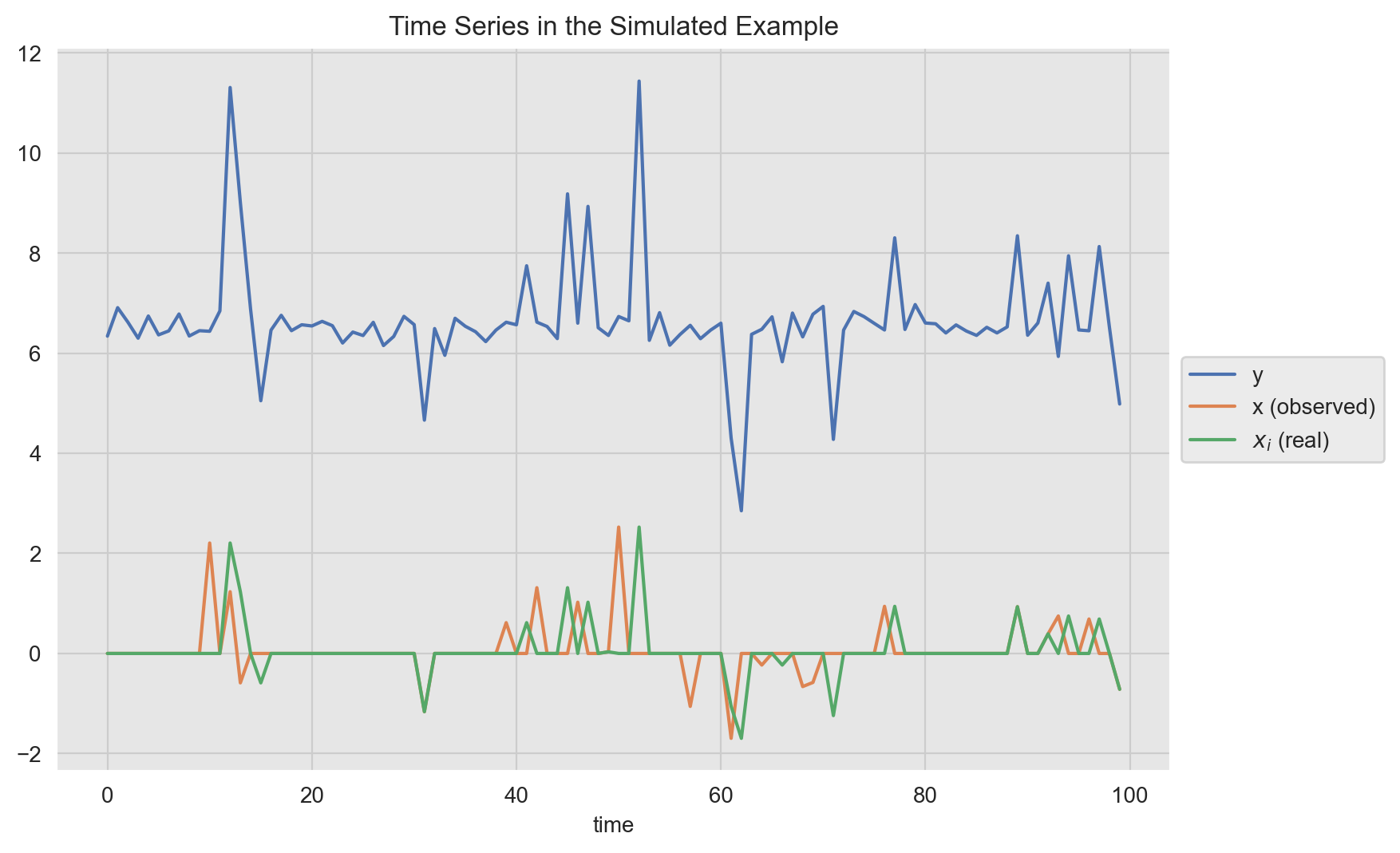}
\caption{Simulated Example Data}
\label{fig:simulated_example}
\end{figure}
\end{center}

The following section demonstrates a simulated univariate example of TVS Regression. Figure \ref{fig:simulated_example} shows the simulated time series. The full code can be found via \href{https://github.com/aaron1rcl/tvs_regression/blob/master/notebooks/1_univariate_example.ipynb}{TVS Regression on GitHub}. 
The input signal has $20$ non-zero impulses which have been drawn from the standard normal distribution. The system is 'off' when $x(t) =0$. In other words, when $x(t) = 0$, $f(x(t)) = 0$. The true shift distribution $\mathcal{T}$ is given by $\mathcal{T} \sim Poisson(\lambda = 2)$. There is one value of $\tau$ for each of the 20 impulses. The green line in Figure \ref{fig:simulated_example} represents the shifted series, corresponding to the time at which the effect occurs. The blue line is the output $y$ which includes a small amount of gaussian noise and an intercept. The values for the parameters were arbitrarily selected. The output $y$ is defined by the following equation:

\begin{align*}
    y = 2(1^TX(t + \mathcal{T})) + 6.5
\end{align*}

\begin{center}
\begin{figure}
\includegraphics[scale=0.5]{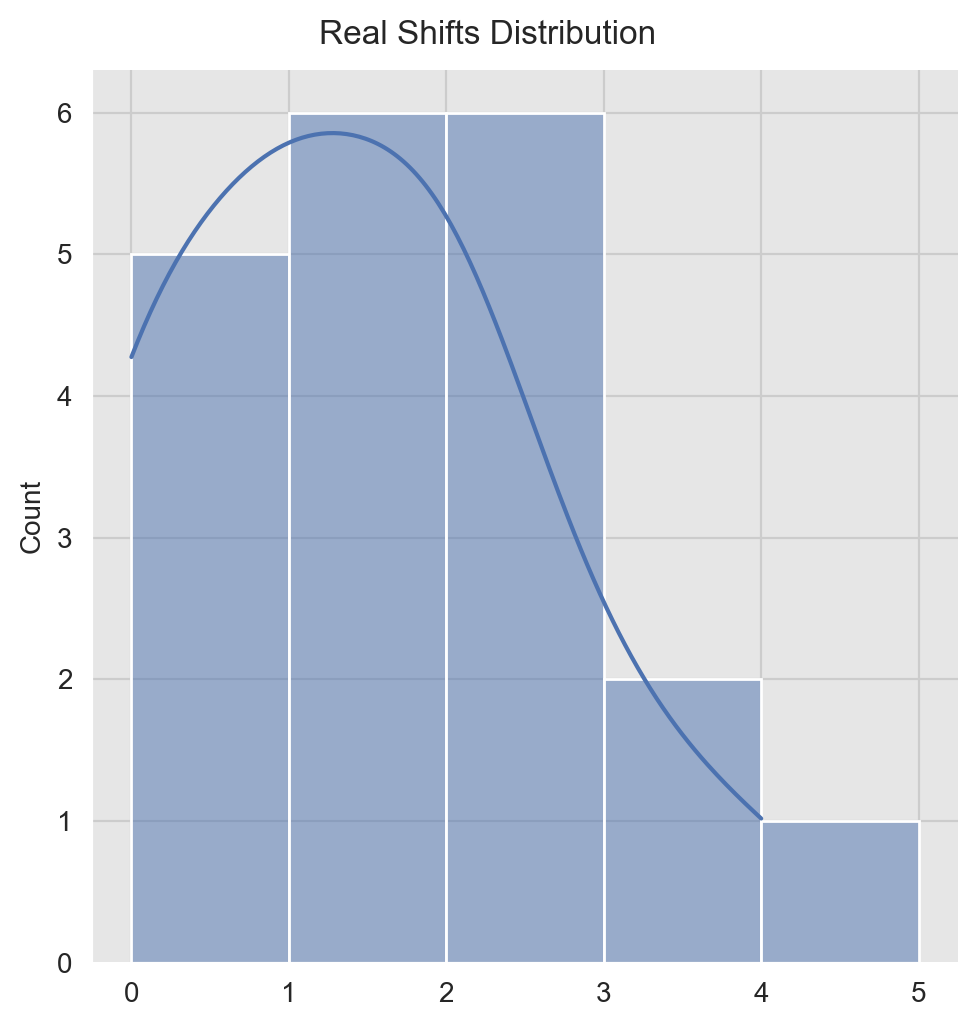}
\caption{Real Shift Distribution $\mathcal{T}$.}
\label{fig:real_shifts}
\end{figure}
\end{center}

A histogram of the actual distribution of $\mathcal{T}$ is shown in Figure \ref{fig:real_shifts}.

\begin{center}
\begin{figure}
\includegraphics[scale=0.6]{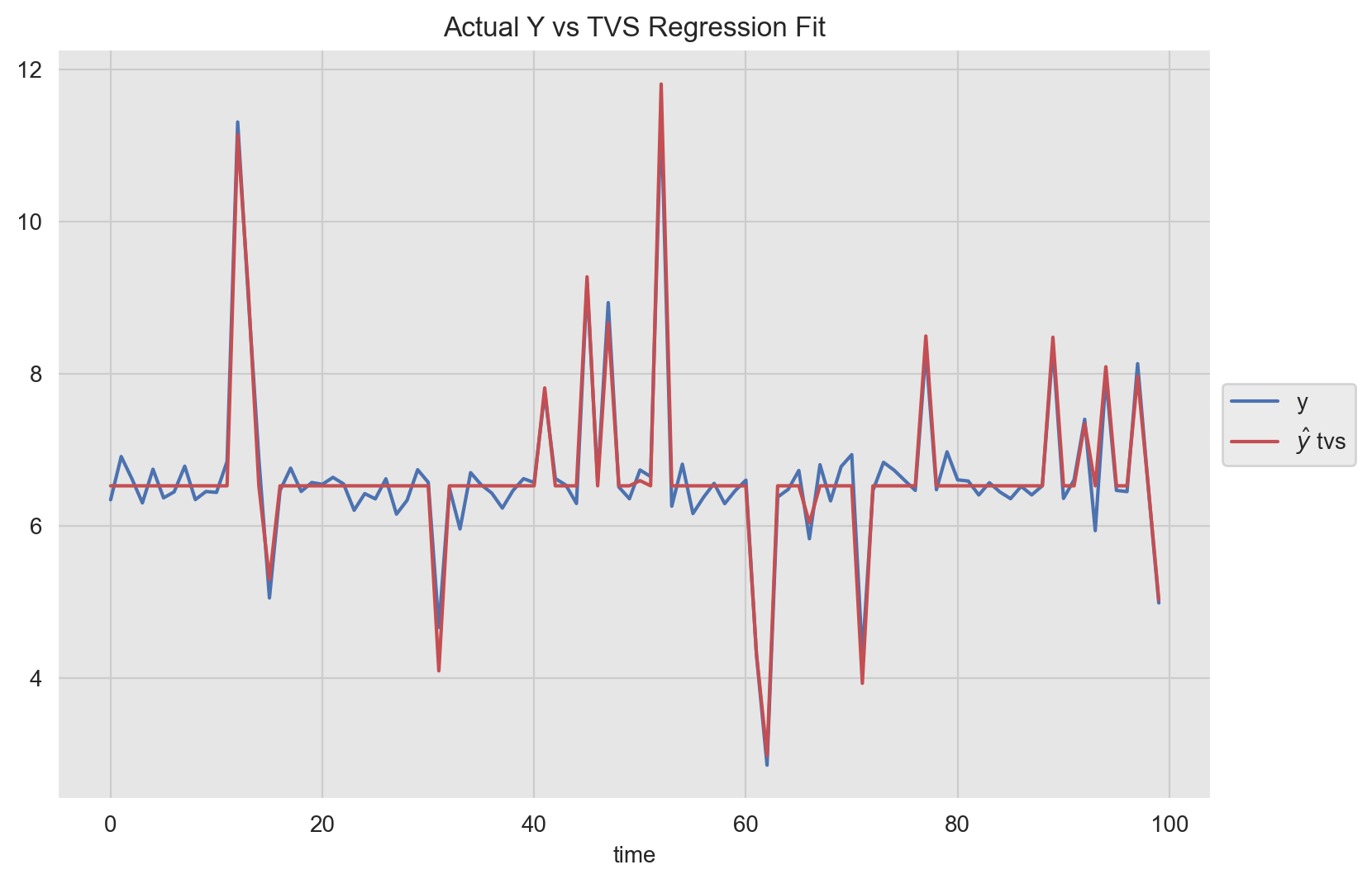} 
\includegraphics[scale=0.3]{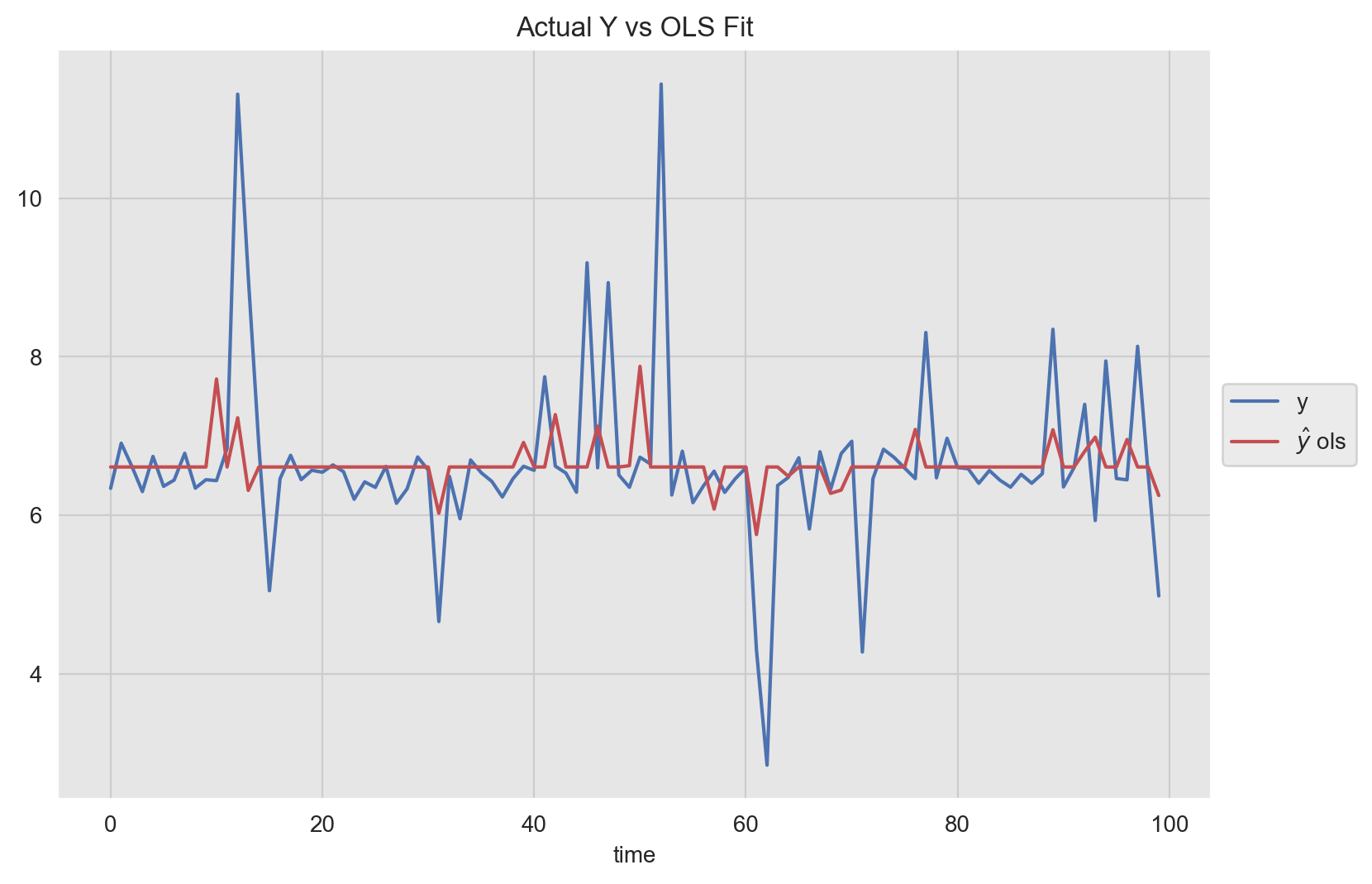} 
\includegraphics[scale=0.3]{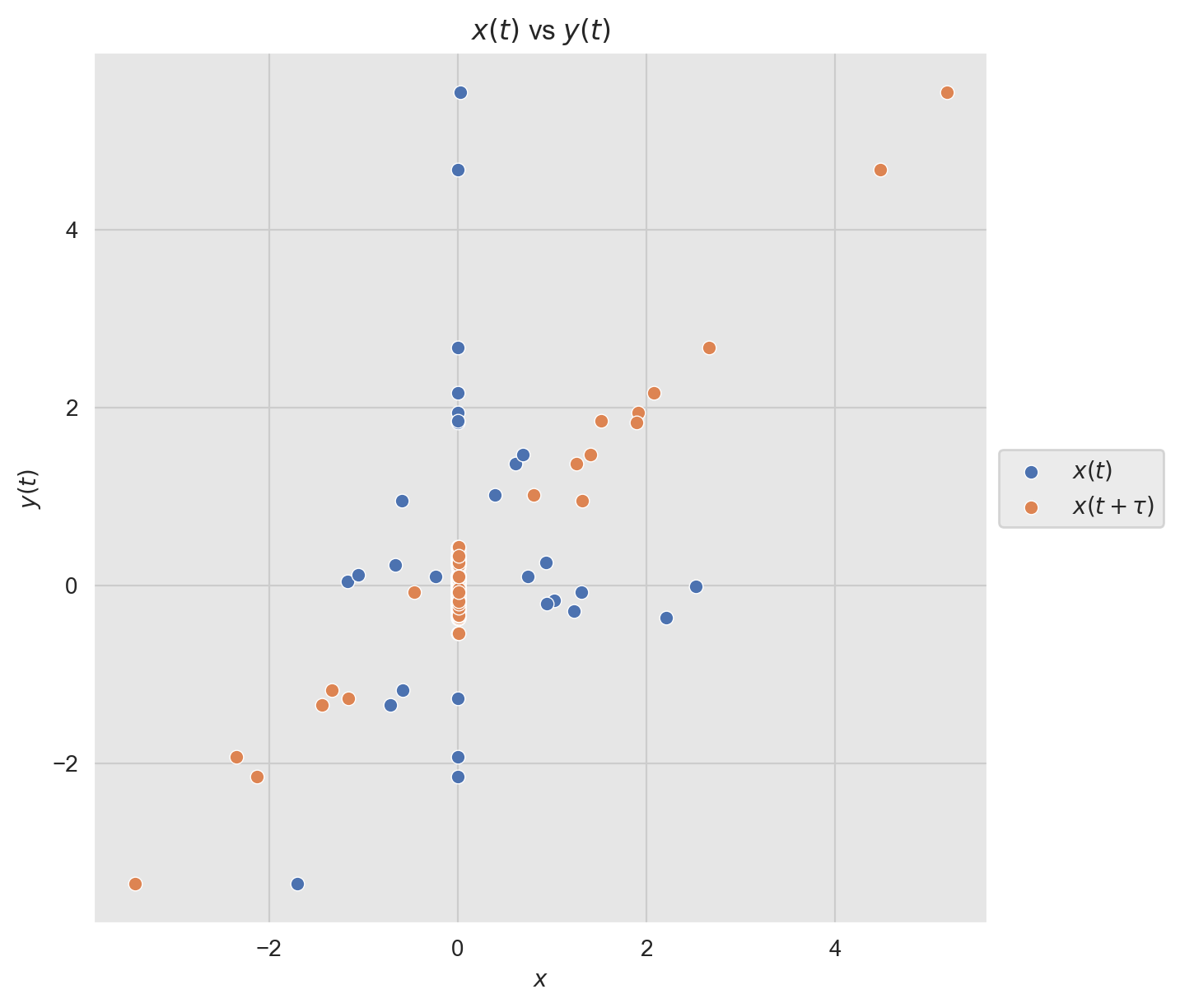} 
\caption{TVS vs OLS model fit.}
\label{fig:tvs_ols_fit}
\end{figure}
\end{center}

\begin{center}
\begin{figure}
\includegraphics[scale=0.5]{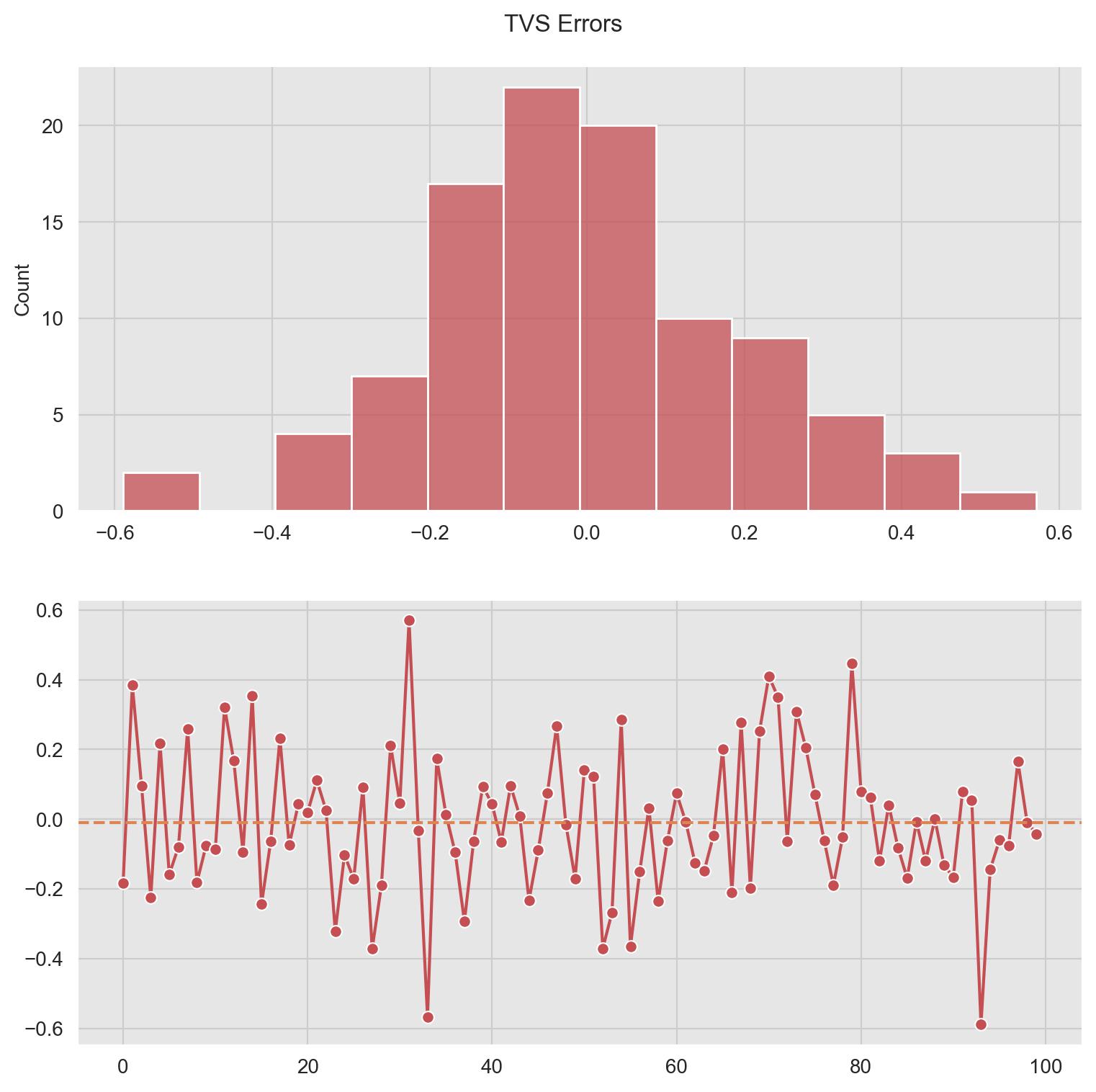}
\caption{Fit Residuals}
\label{fig:tvs_errors}
\end{figure}
\end{center}

\begin{center}
\begin{figure}
\includegraphics[scale=0.6]{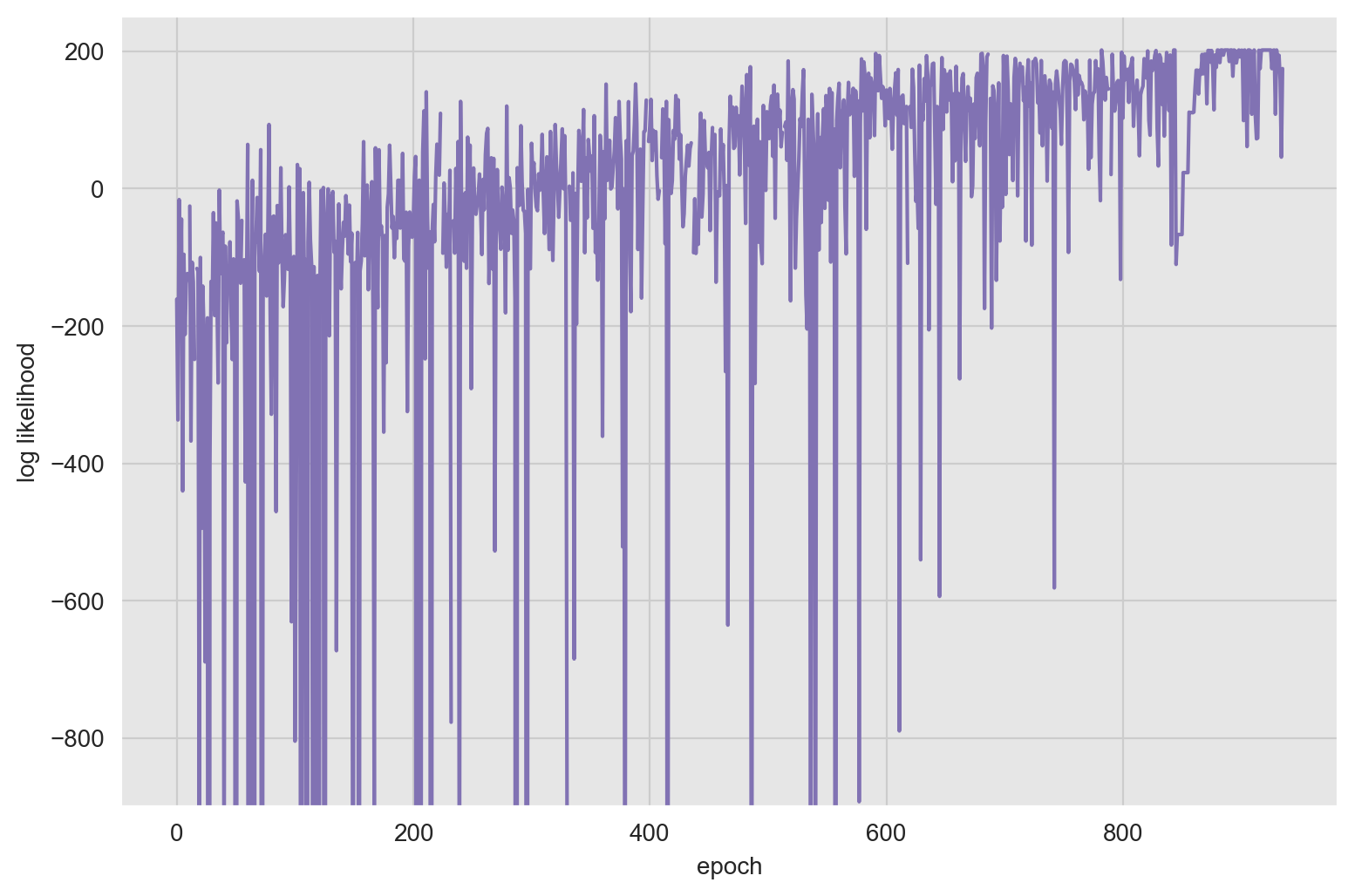} 
\caption{Convergence}
\label{fig:convergence}
\end{figure}
\end{center}

After fitting the model, we obtain the following results. Figure \ref{fig:tvs_ols_fit} shows the model fit (denoted TVS) and a comparison with standard OLS. Figures \ref{fig:tvs_errors} and Figure \ref{fig:convergence} show the error distribution of the TVS fit and the convergence of the TVS model respectively.

The true values of $\mathcal{T}$ (true shifts) and the estimated values $\mathcal{T}_{est}$ are:

\begin{align*}
\mathcal{T} &= 
\left(
\begin{array}{cccccccccccccccccccc}
2 & 1 & 2 & 0 & 2 & 3 & 1 & 0 & 2 & 4 & 1 & 2 & 3 & 2 & 1 & 0  & 0 & 1 & 1 & 0 
\end{array}  
\right) \\
\mathcal{T}_{est} &= 
\left(
\begin{array}{cccccccccccccccccccc}   
2 & 1 & 2 & 0 & 2 & 3 & 1 & 1 & 2 & 4 & 1 & 2 & 3 & 2 & 1 & 0  & 0 & 1 & 1 & 0 
\end{array}  
\right)
\end{align*}

\begin{center}
\begin{table}
\begin{tabular}{|c|c|c|c|}
\hline
Parameter & TVS Regression & OLS Regression & True Value \\
\hline
$\beta$ & 2.09 & 0.50 & 2.00  \\
\hline
$Intercept$ & 6.53 & 6.62 & 6.50 \\
\hline
$\lambda_{\tau}$ & 1.54 & $N/A$ & 1.40 \\
\hline
$\sigma_{\epsilon}$ & 0.20 & 1.03 & 0.20 \\
\hline
\end{tabular}
\caption{ Parameter comparison between TVS and OLS regression against the true value.} 
\label{results_table}
\end{table}
\end{center}

As shown in Table \ref{results_table}, the estimated values for $\beta$ and $\sigma_{\varepsilon}$ are significantly improved by taking into account the stochastic time delay noise. The estimated $\beta$ using OLS regression is $0.5$, compared to the true value $2$. Even a small amount of noise in the time axis causes attenuation limiting the usefulness of OLS for these problems. In comparison, the TVS Regression algorithm estimates $\beta$ to be $2.09$, much closer to the true value.\\

While the example is based on simulated data, we believe that the experiment demonstrates clear potential for improvement in the modelling of real world systems with stochastic time delay noise.

\section{Conclusion}

We have proposed a form of regression analysis suited to the modelling of stochastic time delay problems. In addition, we have shown the feasibility of the approach and its performance on simulated data. Our approach allows for consistently improved estimation and prediction when the input is affected by noise in the time domain. To our knowledge, the method is novel for this class of problem.\\

We propose two extensions as future work. The first is to extend the method to multiple regression. We believe that the extension can be built on the same fundamental ideas presented in this document. Next, the model could be extended to include distributed lag structures. A distributed lag structure is where past values of the impulse influence future values of the output \cite{almon}. Through these extensions we can begin to tackle a number of practical problems that are defined by stochastic relationships between the input and output e.g. the blood glucose problem described in the introduction. In addition, we could also begin to apply the model to forecasting problems where the inputs are affected by stochastic time delay.

\bibliographystyle{acm}
\bibliography{references} 

\begin{thebibliography}{10}

\bibitem{almon}
{\sc Almon, S.}
\newblock The distributed lag between capital appropriations and net
  expenditures,.
\newblock {\em Econometrica}, 33 (1965), 178--196.

\bibitem{Amornbunchornvej_2019}
{\sc Amornbunchornvej, C., Zheleva, E., and Berger-Wolf, T.~Y.}
\newblock Variable-lag granger causality for time series analysis.
\newblock {\em 2019 IEEE International Conference on Data Science and Advanced
  Analytics (DSAA)\/} (Oct 2019).

\bibitem{ml}
{\sc Bianchi, F.~M., Maiorino, E., Kampffmeyer, M., Rizzi, A., and Jenssen, R.}
\newblock {\em Recurrent Neural Networks for Short-Term Load Forecasting: An
  Overview and Comparative Analysis}.
\newblock 01 2017.

\bibitem{dtlr}
{\sc Chandorkar, M., Furtlehner, C., Poduval, B., Camporeale, E., and Sebag,
  M.}
\newblock Dynamictime lag regression: Predicting what and when.
\newblock {\em ICLR 2020 - 8th International Conference onLearning
  Representations\/} (April 2020).

\bibitem{deming}
{\sc Deming, W.}
\newblock The application of least squares.
\newblock {\em Philos. Mag Ser. 7 11\/} (1931), 146--158.

\bibitem{granger}
{\sc Granger, C.}
\newblock Investigating causal relations by econometric models and
  cross-spectral methods.
\newblock {\em Econometrica}, 37 (1969), 424--438.

\bibitem{dtw}
{\sc Müller, M.}
\newblock {\em Dynamic Time Warping. Information Retrieval for Music and
  Motion}.
\newblock Springer, 2007.

\bibitem{dynr}
{\sc Ou, L., Hunter, M.~D., and Chow, S.-M.}
\newblock What's for {dynr}: A package for linear and nonlinear dynamic
  modeling in r.
\newblock {\em The R Journal 11\/} (2019), 1--20.

\bibitem{spearman}
{\sc Spearman, C.}
\newblock The proof and measurement of association between two things.
\newblock {\em American Journal of Psychology}, 15 (1904), 72–101.

\bibitem{diff_ev}
{\sc Storn, R., and Price}.
\newblock Differential evolution - a simple and efficient heuristic for global
  optimization over continuous spaces.
\newblock {\em Journal of Global Optimization 11\/} (1997).

\bibitem{eiv_sysid}
{\sc Söderström, T.}
\newblock {\em Errors-in-Variables Methods in System Identification}.
\newblock Springer, 2018.

\bibitem{scipy}
{\sc Virtanen, P., Gommers, R., Oliphant, T.~E., Haberland, M., Reddy, T.,
  Cournapeau, D., Burovski, E., Peterson, P., Weckesser, W., Bright, J., {van
  der Walt}, S.~J., Brett, M., Wilson, J., Millman, K.~J., Mayorov, N., Nelson,
  A. R.~J., Jones, E., Kern, R., Larson, E., Carey, C.~J., Polat, {\.I}., Feng,
  Y., Moore, E.~W., {VanderPlas}, J., Laxalde, D., Perktold, J., Cimrman, R.,
  Henriksen, I., Quintero, E.~A., Harris, C.~R., Archibald, A.~M., Ribeiro,
  A.~H., Pedregosa, F., {van Mulbregt}, P., and {SciPy 1.0 Contributors}}.
\newblock {{SciPy} 1.0: Fundamental Algorithms for Scientific Computing in
  Python}.
\newblock {\em Nature Methods 17\/} (2020), 261--272.

\bibitem{blood_glucose}
{\sc Woldaregay, A.~Z., Årsand, E., Walderhaug, S., Albers, D., Mamykina, L.,
  Botsis, T., and Hartvigsen, G.}
\newblock Data-driven modeling and prediction of blood glucose dynamics:
  Machine learning applications in type 1 diabetes.
\newblock {\em Artificial Intelligence in Medicine 98\/} (2019), 109--134.

\bibitem{deep_transformers}
{\sc Wu, N., Green, B., Ben, X., and O'Banion, S.}
\newblock Deep transformer models for time series forecasting: The influenza
  prevalence case, 2020.

\end{thebibliography}

\end{document}